# Characterizing the Role of Complex Power in Small-Signal Synchronization Stability of Multi-Converter Power Systems

Fuyilong Ma, Huanhai Xin$^*$, Zhiyi Li, Linbin Huang

*Abstract*—Small-signal synchronization instability (SSI) may be triggered when a grid-connected converter is operated in weak grids. This problem is highly related to the active and reactive power (referred to as complex power) generation or consumption of the converter. Such an instability phenomenon manifests as power oscillations within the bandwidth frequency of phase-locked loop (PLL). However, in a multi-converter power system (MCPS), it is challenging to characterize the role of each converter's complex power in SSI issues due to the high dimension of system dynamics. In this paper, we focus on the impact regularity of complex power on the small-signal synchronization stability of the MCPS. Firstly, we formulate the model of the MCPS and develop a systematical stability indicator to explicitly quantify the stability margin under various operating complex power. Then, based on the proposed stability indicator, the "resultant force" of complex power on the stability margin of the MCPS can be analytically characterized into a weighted summation form. Furthermore, we elaborate on the impact of each converter's active power and reactive power, respectively. Notably, we demonstrate the cancellation effect between the active power generation and consumption on the stability margin, which ultimately helps prevent SSI issues in the MCPS. The correctness of theoretical findings is well verified by simulations results.

## I. Introduction

UNDER the drive to decarbonize electrical energy, renewable energies have been rapidly developed and used in the modern power systems [1]. To facilitate the seamless integration of large-scale renewable energy such as wind and photovoltaic power into the power systems, power electronic converters have been widely adopted due to their flexible control over power generation and consumption [2]. As this trend continues, the system is evolving into the multi-converter power system (MCPS), where conventional synchronous generators (SGs) are being gradually replaced by converters [3]. Unlike SGs, these converter-based apparatuses lack physical rotors governing their synchronization in the ac grid; instead, they rely on digital control algorithms such as phase-locked loop (PLL) [5]. This shift has significantly brought about changes in the dynamic performance of power systems and has even affected the small-signal synchronization stability of the system [5],[6]. For instance, sustained oscillations have been observed in realistic wind farms in Xinjiang (China) [7]. These oscillations are identified as the small-signal synchronization instability (SSI) issue and pose a serious threat to the acceptance of renewable resources for the grid [8]. Therefore, ensuring the small-signal synchronization stability of converters is a critical concern for today's power systems.

One of the key factors affecting the small-signal synchronization stability is the operating power points of converters. Converter-interfaced apparatuses typically operate at various points, involving both active and reactive power (referred to as complex power) generation or consumption [9]-[10]. It has been observed that SSI issues can arise under specific operating power points due to the interaction between the fast-acting PLL control and the power network [11]. This instability phenomenon manifests as power oscillations within the PLL's bandwidth frequency [12]. Notably, large active power transfers between the converter and the rest of the system (such as a weak grid) can more readily trigger SSI issues [8]. Moreover, identifying the operating power points of MCPS that pose a high SSI risk remains a significant concern for grid operators [7]. With multiple converters in the system, each containing tens of states [13] and the wide complex power injection range [10], the dimension of system dynamics becomes very high. In light of this, it is challenging to characterize the impact of operating power points variations on small-signal synchronization stability. Also, ensuring a safe distant from high-risk operating power points is an equally demanding task.

For a single-converter grid-connected system, the impact of operating power points on the small-signal synchronization stability has been comprehensively studied [12]-[16]. By employing the classic eigenvalue analysis, References [14], [15] investigate light and heavy load conditions of WTGs effected by the wind speed and demonstrate that the high active power export is the one of root causes of SSI. Moreover, the effects of power factor and reactive power are examined in [16] based on the popular impedance/admittance spectrum analysis. The SSI mechanism is introduced as the series resonance of the equivalent circuit when the power factor of the converter can be considered as a part of impedance elements [16]. However, these methods do not provide the solution to tackle the interactions among multiple converters [17], and hence difficult to disclose the impact of operating power points on SSI issues

F. Ma, H. Xin and Z. Li are with the College of Electrical Engineering, Zhejiang University, Hangzhou 310027, China. (e-mails: 12210001@zju.edu.cn; xinhh@zju.edu.cn; zhiyi@zju.edu.cn). (*Corresponding author*: Huanhai Xin.)

L. Huang is with the Department of Information Technology and Electrical Engineering, ETH Zürich, 8092 Zürich, Switzerland. (e-mail: linhuang@ethz.ch).

in the MCPS.

Recently, there have been a few efforts on the small-signal synchronization stability of the MCPS with operating power points varying. Impedance-based analysis methods in MCPSs include the frequency-scanning technique to screen for SSI issue risks under specific operating points [17], or over a very narrow operating range [18]. To monitor wider possible operating range, the security region or stability region methods are developed based on the aggregated impedance modeling of the MCPS [19]-[20]. The frequency and damping of SSI modes for various operating points can be calculated by the discrete numerical technology, and thus the stable area for a full range of operating points can be characterized [20]. Nevertheless, these methods implicitly trace the stability margin of MCPS effected by operating power points varying on the basis of numerical tests, rather than the analytical method, so they are difficult to relate SSI issues back to the physical mechanism.

Overall, the current research in the MCPS still reveals the following gaps. The first gap lies in assessing the small-signal synchronization stability margin under various operating power points of converters based on analytical indicators with physical insights. The second gap involves understanding the impact regularity of operating power points varying on the stability, while considering the intricate interplay of converters' complex power. To fill these gaps, we propose an analytic approach to explicitly quantify the stability margin under various operating power points. Additionally, our proposed stability indicator sheds light on the interactive impact of multi-converter complex power, akin to a "resultant force" on the system stability. The key contributions are threefold:

1) In the context of the MCPS, analyzing small-signal synchronization stability becomes intricate due to interactions between the power network and multiple converters. To reduce the complexity, we decouple the MCPS model into a set of simple subsystems and strictly ensure the equivalence of the small-signal synchronization stability condition before and after decoupling.

2) A systematical stability indicator for small-signal synchronization stability of MCPS is developed, allowing for analytically reflect the system stability under various operating power points. Instead of relying on cumbersome numerical tests, we assess the stability margin by analogously applying the positive-net-damping coefficient with physical insights into the synchronization of MCPS.

3) Based on a weighted summation form of the proposed stability indicator, we quantitively clarify the sensitivity regularity of converter's complex power variations with respect to the system stability. Additionally, we uncover the cancellation effect between the active power generation and consumption, which helps prevent SSI issues in the MCPS.

The rest of this paper is organized as follows: Section II introduces the modeling of MCPS with PLL dynamics; Section III proposes the small-signal synchronization stability indicator in the MCPS; Section IV investigates the role of complex power on the stability margin; Section V provides simulation results; Section VI concludes the paper.

## II. Modeling of Multi-converter Power System

To pave the way for the stability analysis in MCPS, we will develop a frequency-domain (i.e., $s$-domain) admittance modeling of power networks and converters to describe small-signal dynamics of the MCPS in this section.

### A. System Descriptions

We consider a MCPS containing $n$ converters (connected to node 1~$n$) with different complex power (i.e., active power and reactive power) generation and consumption in the networks, as depicted in Fig.1. In general, the MCPS can be widely found in renewable energy transmission power systems with interconnected converter-interfaced apparatuses such as wind turbine generators (WTGs), and energy storages (ESs) [2]. Without loss of generality, the most commonly-used PLL-based control scheme is considered for the converters [13], as shown in Fig.2. The control scheme in Fig.2 comprises the PLL, the inner current controller (ICC), the active power controller (APC) and reactive power controller (RPC). The operating complex power of converters in Fig.1 can be flexibly adjusted under such control scheme in Fig.2.

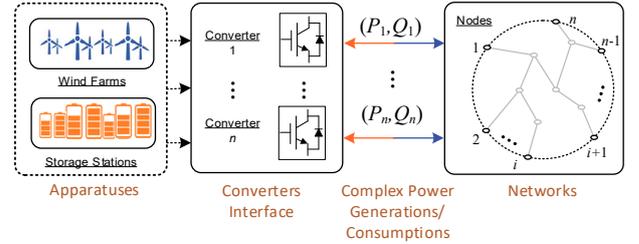

Fig.1 Diagram of multi-converter power system with complex power generation and consumption.

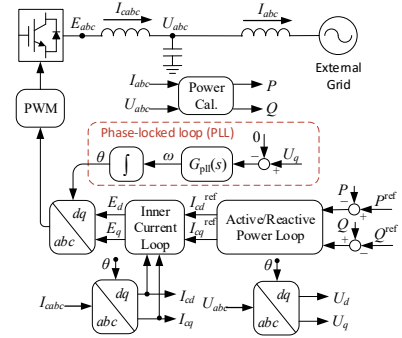

Fig.2 Control diagram of a PLL-based converter.

Given that the converter system in Fig.2 is designed in the local $dq$ frame, Fig. 3 (a) illustrates the relationship between the global $xy$ frame (i.e., the angular frequency of $xy$ frame is the rated synchronous angular frequency $\omega_0 = 100\pi$ rad/s) and the local $dq$ frame (i.e., the angular frequency of $dq$ frame $\omega$ is generated by PLL) [16]. The angle between the local $dq$ frame and the global $xy$ frame $\theta$ is governed by the PLL's equations in Fig.2 as follow

$$\theta = \frac{\omega}{s} = \frac{1}{s} G_{\text{pll}}(s) U_q \qquad (1)$$

where $G_{\text{pll}}(s) = k_{\text{ppll}} + k_{\text{ipll}}/s$ represents the transfer functions of the PLL, with $k_{\text{ppll}}$ and $k_{\text{ipll}}$ being the proportion and integral coefficients of the PI controller of PLL; $U_q$ represents the $q$-axis component of terminal voltage of a converter.

Furthermore, to better illustrate the physical insights of synchronization dynamics in MCPS, the phase domain (i.e., polar frame) is introduced as shown in Fig. 3(b) [16]. Fig. 3(b) shows the relation between the phasor domain (polar frame) and the global $xy$ frame (the polar frame's polar axis is aligned with $x$ axis of $xy$ reference frame).

The admittance modeling for the MCPS shown in Fig.1 will be developed by following two steps: 1) formulating the admittance models of multiple converters and power networks in the global $xy$ frame; 2) transforming these admittance models from the global $xy$ frame to phase domain and integrating converters' model with the power network model.

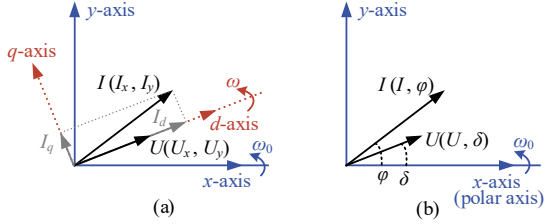

Fig. 3 Reference frameworks in the MCPS. a) controller's $dq$ frame and global system $xy$ frame. (b) phase domain (polar frame) and global system $xy$ frame.

*B. Admittance Modeling in Global XY Frame*

Under the small-signal scenario, the dynamics of MCPS can be captured by the admittance modeling in global $xy$ reference frame. This model can be constructed by the linearized equations of injected current and terminal voltage at the converter and grid terminals, respectively.

1) Power Network Model

At the grid terminals, the linearized equation of the connected power networks can be represented based on the assumptions in [21] as

$$-\Delta \boldsymbol{I}_{xy} = \boldsymbol{Y}_{net}(s)\Delta \boldsymbol{U}_{xy} = \boldsymbol{B} \otimes \begin{bmatrix} \beta(s) & \alpha(s) \\ -\alpha(s) & \beta(s) \end{bmatrix} \Delta \boldsymbol{U}_{xy} \quad (2)$$

where $\Delta(\cdot)$ represents the small-signal variation of the variable; $\Delta \boldsymbol{I}_{xy}=[\Delta \boldsymbol{I}_{xy,1},\ldots,\Delta \boldsymbol{I}_{xy,n}]^T$ is the vector of $\Delta \boldsymbol{I}_{xy,i}=[\Delta I_{x,i},\Delta I_{y,i}]^T$ $i=\{1,\ldots,n\}$, which represents the variation of injected current in the global $xy$ reference frame; $\Delta \boldsymbol{U}_{xy}=[\Delta \boldsymbol{U}_{xy,1},\ldots,\Delta \boldsymbol{U}_{xy,n}]^T$ is the vector of $\Delta \boldsymbol{U}_{xy,i}=[\Delta U_{x,i},\Delta U_{y,i}]^T$ $i=\{1,\ldots,n\}$, which represents the variation of terminal voltage in the global $xy$ reference frame; $\boldsymbol{B}=[B_{ij}] \in \mathbb{R}^{n \times n}$ represents the susceptance matrix of the Kron-reduced networks which reserves $n$ nodes connected to converters; $B_{ij}=B_{ji}=1/(\omega_0 L_{ij})$ and $B_{ii}=-\sum_{j=1,j\neq i}^n B_{ij}$ with $L_{ij}$ being the line susceptance. $\beta(s) = s\omega_0/(s^2+\omega_0^2)$ and $\alpha(s) = \omega_0^2/(s^2+\omega_0^2)$; $\otimes$ denotes the Kronecker product. It has shown that the error introduced by (2) is very small in [21] so that it provides a simple and sufficiently precise model of networks for analysis.

2) Converters Model

At each converter terminal, the linearized equation of terminal voltage and current is commonly represented in the local $dq$ reference frame as

$$\Delta \boldsymbol{I}_{dq,i} = \boldsymbol{Y}_{dq,i}(s)\Delta \boldsymbol{U}_{dq,i}, i=\{1,\ldots,n\} \quad (3)$$

where $\Delta \boldsymbol{I}_{dq,i}=[\Delta I_{d,i},\Delta I_{q,i}]^T$ and $\Delta \boldsymbol{U}_{dq,i}=[\Delta U_{d,i},\Delta U_{q,i}]^T$ represent the vectors of the injected current and terminal voltage of the $i$-th converter in $d$-axis and $q$-axis frame. $\boldsymbol{Y}_{dq,i}(s)$ represents the admittance model for the converter in local $dq$ frame, which contains the dynamics of converter's controllers such as ICC, APC and RPC. Please refer to [22] for a detailed derivation and expression of $\boldsymbol{Y}_{dq,i}(s)$.

To facilitate the integration of converter model with the power network model to construct the model for the system in Fig.1, the components of the injected current and terminal voltage for each converter represented in the local $dq$ frame will be transformed into the global $xy$ frame, by using (4)-(5)

$$\Delta \boldsymbol{I}_{dq,i} = \Delta \boldsymbol{I}_{xy,i} + \boldsymbol{I}_{0i}\Delta \theta_i \quad (4)$$

$$\Delta \boldsymbol{U}_{dq,i} = \Delta \boldsymbol{U}_{dq,i} + \boldsymbol{U}_{0i}\Delta \theta_i \quad (5)$$

where $\theta_i$ represents the angle between the local $dq$ frame and the global $xy$ frame governed by the $i$-th converter's PLL; $\boldsymbol{U}_{0i}=[U_{qi},-U_{di}]^T=-[0,U_i]^T$ represents the vector of the $i$-th converter's terminal voltage in $d$-axis and $q$-axis frame at the steady state; $U_i$ represents the steady amplitude of the $i$-th converter terminal voltage; $\boldsymbol{I}_{0i}=[I_{qi},-I_{di}]^T=[Q_i/U_i, P_i/U_i]^T$ represents the vector of the $i$-th converter's injected current in d-axis and q-axis frame at the steady state; $P_i=U_i I_{di}$ and $Q_i=-U_i I_{qi}$ represents the operating active power and reactive power of the $i$-th converter, respectively; the positive direction of complex power is specified from the converter terminal to the networks.

Substituting (4) and (5) into (3) and linearizing (1) yield

$$\Delta \boldsymbol{I}_{xy,i} = \boldsymbol{Y}_{con,i}(s,P_i,Q_i)\Delta \boldsymbol{U}_{xy,i} \quad (6)$$

where $\boldsymbol{Y}_{con,i}(s,P_i,Q_i)$ represents the admittance model for the $i$-th converter in global $xy$ frame containing the operating power points $P_i$ and $Q_i$. For the detailed derivation and expression of $\boldsymbol{Y}_{con,i}(s,P_i,Q_i)$, we refer to [22].

*C. Admittance Modeling in Phasor Domain*

To facilitate the analysis of the synchronization stability with physical insights, we convert the admittance model for power networks and converters from the global $xy$ frame into the phasor domain.

According to the relationship between the phasor domain and the global $xy$ frame shown in Fig. 3(b), the vector (voltage or current) under the disturbance in the global $xy$ frame can be transformed into the phasor domain as

$$\begin{bmatrix} \Delta U_{x,i} \\ \Delta U_{y,i} \end{bmatrix} = \begin{bmatrix} \cos\delta_i & -\sin\delta_i \\ \sin\delta_i & \cos\delta_i \end{bmatrix} \begin{bmatrix} \Delta U_i \\ U_i\Delta\delta_i \end{bmatrix} \approx \begin{bmatrix} \Delta U_i \\ U_i\Delta\delta_i \end{bmatrix} \quad (7)$$

$$\begin{bmatrix} \Delta I_{x,i} \\ \Delta I_{y,i} \end{bmatrix} = \begin{bmatrix} \cos\varphi_i & -\sin\varphi_i \\ \sin\varphi_i & \cos\varphi_i \end{bmatrix} \begin{bmatrix} \Delta I_i \\ I_i\Delta\varphi_i \end{bmatrix} =: \boldsymbol{T}_{\varphi,i} \begin{bmatrix} \Delta I_i \\ I_i\Delta\varphi_i \end{bmatrix} \quad (8)$$

where $I_i$ and $\varphi_i$ represent the amplitude and angle of the injected current of the $i$-th converter $i=\{1,\ldots,n\}$; $\delta_i$ represent the angle of the terminal voltage of the $i$-th converter, and we consider $\cos\delta_i \approx 1, \sin\delta_i \approx 0$ as assumed in [21].

By substituting (7)-(8) into (2), the power network model in the phasor domain can be developed as

$$\Delta \boldsymbol{U}_\delta = -\boldsymbol{Y}_{net}^{-1}(s)\boldsymbol{T}_\varphi^{-1}\Delta \boldsymbol{I}_\varphi \quad (9)$$

Where $\Delta \boldsymbol{U}_\delta=[\Delta \boldsymbol{U}_{\delta,1},\ldots,\Delta \boldsymbol{U}_{\delta,n}]^T$ is the vector of $\Delta \boldsymbol{U}_{\delta,i}=[\Delta U_i, U_i\Delta\delta_i]^T$, which represents the variation of terminal voltage in the phasor domain; $\Delta \boldsymbol{I}_\varphi=[\Delta \boldsymbol{I}_{\varphi,1},\ldots,\Delta \boldsymbol{I}_{\varphi,n}]^T$ is the vector of $\Delta \boldsymbol{I}_{\varphi,i}=[\Delta I_i, I_i\Delta\varphi_i]^T$, which represents of the variation of injected current in the phasor domain; $\boldsymbol{T}_\varphi = \text{diag}\{\boldsymbol{T}_{\varphi i}\}$ represents the block-diagonal matrix whose elements are defined in (8).

Similarly, the admittance model of converters in the phase domain can be developed as

$$\Delta \boldsymbol{I}_\varphi = \boldsymbol{T}_\varphi diag\{Y_{con,i}(s, P_i, Q_i)\}\Delta \boldsymbol{U}_\delta \qquad (10)$$

By integrating converters' model in (10) with the power network model in (9), the closed-loop system dynamics of the MCPS can be illustrated in Fig. 4. Note that the additive disturbances in the voltage under the phasor domain can be considered in the closed-loop feedback system.

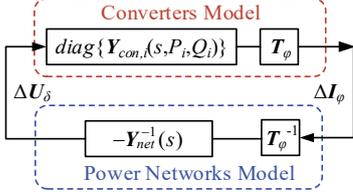

Fig. 4 Closed-loop diagram of the MCPS.

## III. SMALL-SIGNAL SYNCHRONIZATION STABILITY ASSESSMENT OF MCPS

In this section, we decouple the modeling of the MCPS into a set of simple subsystems in terms of small-signal synchronous stability assessment. Also, a stability indicator is proposed based on the analogy application of the positive-net-damping stability criterion in the MCPS.

### A. Positive-net-damping Stability Criterion Basic

The positive-net-damping stability criterion was initially introduced for turbine-generator torsional interaction analysis of the SG-dominated system [23],[24]. The interaction mechanism can be physically explained as the SG rotor motion driven by the mechanical and electrical torque. To permit stability analysis, the electrical and mechanical dynamics under the small-signal scenario can be described by the frequency functions $G_e(j\omega)$ and $G_m(j\omega)$, respectively, and be formed as a closed-loop system, as shown in Fig. 5. Fig. 5 shows that the dynamics of rotor speed $\omega_r$ as the output is governed by the mechanical and electrical torque (i.e., $T_m$ and $T_e$). The frequency functions $G_e(j\omega)$ and $G_m(j\omega)$ on the control loop can be expressed as

$$G_e(j\omega) = D_e(\omega) + jK_e(\omega) \qquad (11)$$

$$G_m^{-1}(j\omega) = D_m(\omega) + jK_m(\omega) \qquad (12)$$

where $D_e(\omega) = \text{Re}[G_e(j\omega)]$ and $D_m(\omega) = \text{Re}[G_m(j\omega)]$ represent the real part of $G_e(j\omega)$ and $G_m^{-1}(j\omega)$, that is, are referred to as the electrical and mechanical damping coefficients [23], respectively; $K_e(\omega) = \text{Im}[G_e(j\omega)]$ and $K_m(\omega) = \text{Im}[G_m(j\omega)]$ represent the imaginary part of $G_e(j\omega)$ and $G_m^{-1}(j\omega)$, that is, are referred to as the electrical and mechanical spring coefficients [23], respectively.

The stability criterion for the system in Fig. 5 involves the evaluation of the net damping at the oscillation angular frequency [25], that is,

$$\begin{cases} D_e(\omega_c) + D_m(\omega_c) > 0 \Rightarrow \text{Stable} \\ D_e(\omega_c) + D_m(\omega_c) \leq 0 \Rightarrow \text{Unstable} \end{cases} \qquad (13)$$

where $\omega_c = 2\pi f_c$ represents the oscillation angular frequency and satisfies $K_m(\omega_c) + K_e(\omega_c) = 0$.

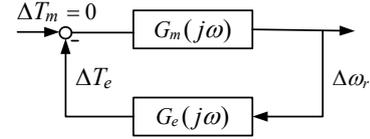

Fig. 5 Illustration of the positive-net-damping stability criterion referred to [24].

### B. Analogy Application in Decoupled MCPS

In this subsection, by analogizing the synchronization of the PLL-based converter driven by the angle of terminal voltage with that of the SG driven by torque, we apply the positive-net-damping stability criterion in the small-signal synchronization stability analysis of MCPS and show how to decouple the MCPS model into a set of simple subsystems.

More specifically, the MCPS model in Fig. 4 can be turned into the format with the PLL generated frequency as the output, like the SG rotor motion illustrated in Fig. 5. To simplify the analysis, we consider that the bandwidth of the PLL is normally narrower than that of ICC and wider than that of APC and RPC [11], [26]. It can be thought that the converter's current feedback is the same with the current references under this situation, that is, $\Delta \boldsymbol{I}_{dq,i} = [\Delta I_{di}, \Delta I_{qi}]^T \approx 0$, $i = \{1, \ldots, n\}$. As a result, we focus on oscillation instability within the bandwidth of PLL in terms of the small-signal synchronization stability.

By substituting (1) into (4) and (8), the injected current of the $i$-th converter can be expressed by its PLL generated frequency

$$\Delta \boldsymbol{I}_{\varphi,i} = -\boldsymbol{T}_{\varphi i} \boldsymbol{I}_{0i} \Delta \theta_i = \boldsymbol{T}_{\varphi i} \frac{\omega_0}{sU_i}\begin{bmatrix} Q_i \\ P_i \end{bmatrix}\Delta \omega_i / \omega_0 \qquad (14)$$

Combing (14) with (10), the admittance modeling of the $i$-th converter in (10) can be deformed as

$$\Delta \omega_i / \omega_0 = \frac{s}{\omega_0} F_{\text{pll},i}(s) U_i \Delta \delta_i := \Gamma_i^{-1}(s)\Delta \delta_i \qquad (15)$$

where $F_{\text{pll},i}(s) = 1/(s/G_{\text{pll},i}(s) + U_i)$ and $G_{\text{pll},i}(s)$ represents the transfer functions of the $i$-th converter's PLL PI controller.

Eq.(15) illustrates how each converter's PLL generates the frequency driven by the angle of the terminal voltage for the synchronization between them. The transfer function $\Gamma_i(s)$ is dominated by the control parameters of PLL. Since we focus on the effect of complex power in MCPS, we assume that all converters share the same PLL parameters to simplify the analysis, that is, $\Gamma_1(s) \approx \ldots \approx \Gamma_n(s) =: \Gamma(s)$. As a result, based on (15), the PLL dynamics of multiple converters can be expressed as

$$\Delta \boldsymbol{\omega}/\omega_0 = \Gamma^{-1}(s)\boldsymbol{I}_n \Delta \boldsymbol{\delta} \qquad (16)$$

where $\Delta \boldsymbol{\omega}/\omega_0 = [\Delta \omega_1, \ldots, \Delta \omega_n]^T/\omega_0$ represents the vector of PLL generated frequency normalized by the rated synchronous angular frequency $\omega_0$; $\Delta \boldsymbol{\delta} = [\Delta \delta_1, \ldots, \Delta \delta_n]^T$ represents the vector of angle of converters' terminal voltage in the phasor domain; $\boldsymbol{I}_n$ denotes the $n$-dimensional identity matrix.

Similarly, by substituting (14) into (9), the admittance modeling of power networks in (9) also can be deformed as

$$\Delta \boldsymbol{\delta} = -(\boldsymbol{B}^{-1}\tilde{\boldsymbol{P}} + \frac{\omega_0}{s}\boldsymbol{B}^{-1}\tilde{\boldsymbol{Q}})\Delta \boldsymbol{\omega}/\omega_0 \qquad (17)$$

where $\tilde{\boldsymbol{P}} = diag\{P_i/U_i^2\}$ and $\tilde{\boldsymbol{Q}} = diag\{Q_i/U_i^2\}$ represent $n$-dimension diagonal matrices whose elements include the operating active power and reactive power of $n$ converters, respectively.

Eq.(17) illustrates that the feedback path from the PLLs generated frequencies $\Delta\boldsymbol{\omega}/\omega_0$ to the angle of the terminal voltage of converters $\Delta\boldsymbol{\delta}$ comprises the network and the complex power of multiple converters.

By substituting $s=j\omega$ in (16) and (17), a closed-loop system can be formed to describe the synchronization of multiple converters based on PLLs, as shown in Fig. 6(a). In Fig. 6(a), $\boldsymbol{G}_{con}(j\omega)$ and $\boldsymbol{G}_{net}(j\omega)$ represent the frequency functions form in (16) and (17), respectively, which are expressed as follows

$$\boldsymbol{G}_{con}^{-1}(j\omega) = \Gamma(j\omega)\boldsymbol{I}_n \quad (18)$$

$$\boldsymbol{G}_{net}(j\omega) = -\boldsymbol{B}^{-1}\tilde{\boldsymbol{P}} + j\omega_R\boldsymbol{B}^{-1}\tilde{\boldsymbol{Q}}, \omega_R = \omega_0/\omega \quad (19)$$

In Fig. 6(a), it is shown that the synchronization of MCPS based on PLLs can be analogous to the format of SG rotor motion in Fig. 5. Recalling the positive-net-damping stability criterion shown in Fig. 5, it is well known that $G_m(j\omega)$ represents how the rotor frequency generated by the mechanical-electrical torque of the SG and the feedback transfer function $G_e(j\omega)$ is determined by the networks. Similarly, in Fig. 6(a), $\boldsymbol{G}_{con}(j\omega)$ represents how the PLLs generate the frequency of each converter driven by its angle of terminal voltage, and the feedback transfer function $\boldsymbol{G}_{net}(j\omega)$ is determined by the coupling networks and complex power. Additionally, although the classic positive-net-damping stability criterion in Fig. 5 is for the single-machine system, the following proposition shows how the MCPS in Fig. 6(a) can be decoupled into simple subsystems and then apply this stability criterion.

**Proposition III.1.** *The n-converter system in Fig. 6(a) can be decoupled into n single-converter subsystems, and the stability of the n-converter system can be characterized by n single-converter subsystems.*

*Proof.* There exists an invertible matrix $\boldsymbol{W}$ to diagonalize the nonsingular matrix $\boldsymbol{G}_{net}(j\omega), \forall \omega > 0$ as

$$\boldsymbol{W}^{-1}\boldsymbol{G}_{net}(j\omega)\boldsymbol{W} = diag\{\lambda_i\{\boldsymbol{G}_{net}(j\omega)\}\} =: \Lambda(j\omega) \quad (20)$$

where $\lambda_i\{\cdot\}$ represents the $i$-th eigenvalue of a matrix.

Consider the following coordinate transformation

$$\Delta\boldsymbol{\omega}_m = \boldsymbol{W}^{-1}\Delta\boldsymbol{\omega}/\omega_0, \Delta\boldsymbol{\delta}_m = \boldsymbol{W}^{-1}\Delta\boldsymbol{\delta} \quad (21)$$

that makes (16) and (17) become

$$\underbrace{\begin{bmatrix}\Delta\omega_{m,1}\\\vdots\\\Delta\omega_{m,n}\end{bmatrix}}_{\Delta\boldsymbol{\omega}_m} = \Gamma^{-1}(j\omega)\boldsymbol{I}_n\underbrace{\begin{bmatrix}\Delta\delta_{m,1}\\\vdots\\\Delta\delta_{m,n}\end{bmatrix}}_{\Delta\boldsymbol{\delta}_m}, \underbrace{\begin{bmatrix}\Delta\delta_{m,1}\\\vdots\\\Delta\delta_{m,n}\end{bmatrix}}_{\Delta\boldsymbol{\delta}_m} = \Lambda(j\omega)\underbrace{\begin{bmatrix}\Delta\omega_{m,1}\\\vdots\\\Delta\omega_{m,n}\end{bmatrix}}_{\Delta\boldsymbol{\omega}_m} \quad (22)$$

Since $\Gamma^{-1}(j\omega)\boldsymbol{I}_n$ and $\Lambda(j\omega)$ are both diagonal matrices in (22), the $n$-converter system in Fig. 6(a) can be intrinsically decoupled as $n$ single-converter subsystems in the modal coordinate, as shown in Fig. 6(b).

Also, since eigenvalues are invariant under the similarity transform, the $n$ single-converter subsystems in Fig. 6(b) have the equivalent characteristic loci [26] with those of the $n$-converter system in Fig. 6(a), that is,

$$\lambda_i\{\boldsymbol{G}_{con}(j\omega)\boldsymbol{G}_{net}(j\omega)\} = \lambda_i\{\boldsymbol{W}\Gamma^{-1}(j\omega)\Lambda(j\omega)\boldsymbol{W}^{-1}\}$$
$$= \lambda_i\{\Gamma^{-1}(j\omega)\Lambda(j\omega)\} \quad (23)$$
$$= \Gamma^{-1}(j\omega)\lambda_i\{\boldsymbol{G}_{net}(j\omega)\}$$

It means that the stability of the system in Fig. 6(a) is determined by $n$ subsystems in (22) based on the generalized Nyquist criterion [27], which concludes the proof ∎.

According to Proposition III.1, applying the positive-net-damping stability criterion in each decoupled subsystem in Fig. 6(b) can be equivalent to the small-signal synchronization stability analysis of the whole system in Fig. 6(a). As a result, according to (13), the MCPS is small-signal synchronization stable if and only if the $i$-th subsystem has the positive net damping for every $i=\{1,\ldots,n\}$, that is,

$$D_{con}(\omega_{ci}) + D_{neti}(\omega_{ci}) > 0, i = \{1,...,n\} \quad (24)$$

where $D_{con}(\omega) = \text{Re}[\Gamma(j\omega)]$ and $D_{neti}(\omega) = \text{Re}[\lambda_i\{\boldsymbol{G}_{net}(j\omega)\}]$ represent the converter-side and network-side damping coefficients of the $i$-th subsystem, respectively; $\omega_{ci}$ represents the oscillation angular frequency of the $i$-th subsystem, which satisfies

$$K_{con}(\omega_{ci}) + K_{neti}(\omega_{ci}) = 0, i = \{1,...,n\} \quad (25)$$

where $K_{con}(\omega) = \text{Im}[\Gamma(j\omega)]$ and $K_{neti}(\omega) = \text{Im}[\lambda_i\{\boldsymbol{G}_{net}(j\omega)\}]$ represent the converter-side and network-side spring coefficients of the $i$-th subsystem, respectively.

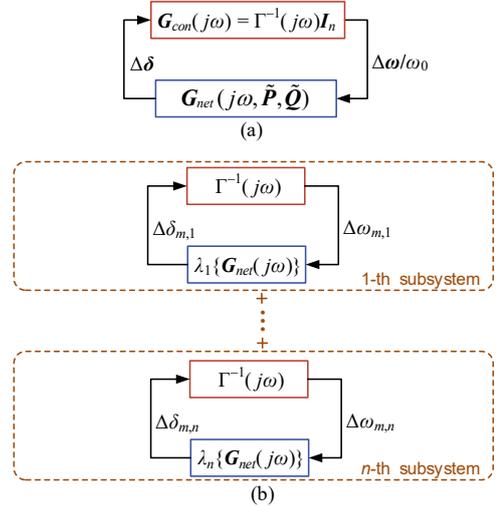

Fig. 6 Decoupling of the MCPS for analyzing small-signal synchronization stability. (a) MCPS model before decoupling; (b) MCPS model after decoupling.

*C. Proposed Stability Indicator*

Based on the above positive-net-damping stability criterion analysis in the MCPS, the subsystem with the minimal net damping should be most concerned, which determines the small-signal synchronization stability margin of the MCPS. For ease of expression, we refer to such a subsystem as the *critical subsystem*. The converter-side and network-side damping coefficients of the critical subsystem can be denoted as $D_{con}(\omega)$ and $D_{net1}(\omega)$.

Notice that the system stability margin can be further determined as the network-side damping coefficient of critical subsystem (hereafter, referred to as the *critical network-side damping coefficient*) $D_{net1}(\omega)$ when the converter-side damping coefficient $D_{con}(\omega)$ is given. This view is based on the fact that converters normally adopt fixed typical control parameters during the grid planning and operating [11]. Hence, we define the small-signal synchronization stability indicator of MCPS as the critical network-side damping coefficient at the oscillation angular frequency, that is,

$$D_{net1} := \text{Re}[\lambda_1\{\boldsymbol{G}_{net}(j\omega_{c1})\}] \quad (26)$$

where $\omega_{c1} = 2\pi f_{c1}$ represents the oscillation angular frequency of the critical subsystem, which satisfies $K_{con}(\omega_{c1}) + K_{net1}(\omega_{c1}) = 0$.

Considering that the complex power of converters are related to the matrix $\boldsymbol{G}_{net}(j\omega_{c1})$ as expressed in (19), the small-signal synchronization stability and stability margin of MCPS at the given operating power points can be identified by using the critical network-side damping coefficient $D_{net1}$ in (26). That is, the MCPS is small-signal synchronization stable if and only if $D_{net1}$ is greater than $-D_{con}(\omega_{c1})$ (i.e. $D_{net1} + D_{con}(j\omega_{c1}) > 0$). Hence, as the proposed stability indicator, the lighter critical network-side damping coefficient can lead to the lighter net damping, and cause the higher small-signal instability risk.

## IV. IMPACT ANALYSIS OF COMPLEX POWER ON SMALL-SIGNAL SYNCHRONIZATION STABILITY

The impact of multi-converter complex power on the small-signal synchronization stability of MCPS can be characterized by the proposed indicator with injected complex power (i.e., critical network-side damping coefficient $D_{net1}$ in (26)). To this end, we first derive the weighted summation form of the proposed indicator to illustrate the contribution of each converter complex power in the system stability. Then, the sensitivity analysis is applied to calculate the effects of active and reactive power variation of each converter. Additionally, we illustrate how the effect of active power generation and consumption in MCPS interact or even cancel each other, and thus improve the small-signal synchronization stability.

### A. Weighted Summation Form with Multi-Converter Complex Power

As the key of small-signal synchronization stability indicator defined in (26), the properties of the eigenvalue $\lambda_1\{\boldsymbol{G}_{net}(j\omega_{c1})\}$ can be illustrated by the following lemmas.

**Lemma 1.** The frequency function matrix $\boldsymbol{G}_{net}(j\omega)$ in (19) is similar to a symmetric matrix as

$$\boldsymbol{G}'_{net}(j\omega) = -\boldsymbol{B}^{-1/2}\tilde{\boldsymbol{P}}\boldsymbol{B}^{-1/2} + j\omega_R \boldsymbol{B}^{-1/2}\tilde{\boldsymbol{Q}}\boldsymbol{B}^{-1/2} \quad (27)$$

*Proof.* The detailed proof is given in Appendix B.

**Lemma 2.** The real and imaginary parts of $\lambda_1\{\boldsymbol{G}_{net}(j\omega_{c1})\}$ can be represented as a weighted summation form with the active power and reactive power of each converter, that is,

$$D_{net1} = \text{Re}[\lambda_1\{\boldsymbol{G}_{net}(j\omega_{c1})\}] = -\sum_{i=1}^{n}\eta_{B1i}(j\omega_{c1})P_i \quad (28)$$

$$K_{net1}(\omega_{c1}) = \text{Im}[\lambda_1\{\boldsymbol{G}_{net}(j\omega_{c1})\}] = \omega_{R1}\sum_{i=1}^{n}\eta_{B1i}(j\omega_{c1})Q_i \quad (29)$$

where $\omega_{R1} = \omega_0/\omega_{c1}$; the $i$-th weighted coefficient represents

$$\eta_{B1i}(j\omega_{c1}) = \phi_{B1i}^2(j\omega_{c1})/U_i^2 \quad (30)$$

where $\phi_{B1i}(j\omega_{c1})$ represents the $i$-th element of the vector $\boldsymbol{\phi}_{B1}(j\omega_{c1}) = \boldsymbol{B}^{-1/2}\boldsymbol{\phi}_1(j\omega_{c1}) \in \mathbb{C}^n$, $i=\{1,...,n\}$; $\boldsymbol{\phi}_1(j\omega_{c1}) \in \mathbb{C}^n$ represents the normalized right eigenvector with respect to the eigenvalue $\lambda_1\{\boldsymbol{G}'_{net}(j\omega_{c1})\}$ of the matrix $\boldsymbol{G}'_{net}(j\omega)$ in (27), which satisfies $\boldsymbol{\phi}_1^*\boldsymbol{\phi}_1 = 1$; $(\cdot)^*$ denotes the conjugate transpose.

*Proof.* The detailed proof is given in Appendix C.

According to Lemma 2, the weighted summation form in (28)-(29) provides the insight into the interactive impact of multi-converter complex power, akin to a "resultant force" on the system stability. The active power of each converter is equipped with the weighted coefficients and summed to the real part of the eigenvalue $\lambda_1\{\boldsymbol{G}_{net}(j\omega_{c1})\}$ (i.e., the stability indicator $D_{net1} = \text{Re}[\lambda_1\{\boldsymbol{G}_{net}(j\omega_{c1})\}]$ defined in (26)); similarly, the reactive power of each converter is weighted and summed to the imaginary part of the eigenvalue $\lambda_1\{\boldsymbol{G}_{net}(j\omega_{c1})\}$ (i.e., $K_{net1}(\omega_{c1}) = \text{Im}[\lambda_1\{\boldsymbol{G}_{net}(j\omega_{c1})\}]$ in (26)). Based on such a form, the impact regularity of complex power on the small-signal synchronization stability of MCPS will be further discussed in the next subsections.

### B. Complex Power Variation Influence

We now analyze how the complex power varying of each converter affects the small-signal synchronization stability in MCPS. From Lemma 2, it can be preliminarily seen that the active power of converters has a direct effect on the stability indicator (i.e., the critical network-side damping coefficient $D_{net1}$ in (28)), while the reactive power of converters has an indirect impact on the stability indicator $D_{net1}$ though the spring coefficient $K_{net1}(\omega_{c1})$ in (29) due to the stability indicator defined at the oscillation angular frequency in (26). The following proposition further illustrates this point.

**Proposition IV.1.** *The active power variations of a converter $P_i$ ($i=\{1,...,n\}$) are sensitive and negatively correlated to the proposed stability indicator $D_{net1}$, and thus significantly change the small-signal synchronous stability margin. While the reactive power variations of a converter $Q_i$ ($i=\{1,...,n\}$) are not sensitive with respect to the proposed stability indicator $D_{net1}$.*

*Proof.* Recall the weighted summation form in (28) in Lemma 2, the sensitivity of the perturbation of each converter's active or reactive power with respect to the proposed stability indicator can be derived as follows

$$\frac{\partial D_{net1}}{\partial P_i} = -\eta_{B1i}(j\omega_{c1}) = -\frac{\phi_{B1i}^2(j\omega_{c1})}{U_i^2} < 0, \frac{\partial D_{net1}}{\partial Q_i} = 0. \quad (31)$$

Thus, based the sensitivity analysis in (31), this concludes the proof ∎.

Proposition 1 demonstrates that increasing the converter's active power can lead to the critical network-side damping coefficient dropping, and thus impairs the small-signal synchronization stability margin. The variation of reactive power is insensitive to the critical network-side damping coefficient, and its indirect impact has been illustrated in (29). Notably, the above result echo the existing feasible engineering experience, that is, limiting the maximum active power effort of converters is an effective means to suppress the oscillations caused by the PLL [7],[8],[15].

Furthermore, it is worth to note that the sensitivity results can identify the most sensitive active power variation of the dominant converter. The norm of weighted coefficients in (31) (i.e., $|\eta_{B1i}(j\omega_{c1})|$, $i=\{1,...,n\}$) can quantify the participation of the $i$-th active power with respect to the stability indicator $D_{net1}$. Hence, the dominant converter with the largest $|\eta_{B1i}(j\omega_{c1})|$ in turn has the most sensitive active power variations for the small-signal synchronization stability.

*C. Interactions Between Power Generation and Consumption*

Based on the weighted summation form of the proposed stability indicator, we now discuss how the interaction between power generation and consumption of converters affects the small-signal synchronization stability in MCPS. We will focus on the interactive effect of active power generation and consumption, since the impact of reactive power variation has been shown to be insensitive to the system stability.

To illustrate the interactive effect of active power generation and consumption, let us consider that there are $k$ converters at active power generation states and $l$ converters with active power consumption states in the MCPS. That is, the positive inertia of the matrix $\tilde{P}$ satisfies $i_+(\tilde{P}) = k$ and the negative inertia has $i_-(\tilde{P}) = l$ ($k+l=n$) [28]. The proposed stability indicator of the MCPS can be represented as

$$D_{net1} = -\sum_{i=1}^{k} \eta_{B1i}(j\omega_{c1})P_{Gi} - \sum_{i=k+1}^{l} \eta_{B1i}(j\omega_{c1})P_{Ci} \quad (32)$$

where $P_{Gi}$ ($i=1,\ldots,k$) represents the generated active power of $k$ converters; $P_{Ci} < 0$ ($i=k+1,\ldots,l$) represents the consumed active power of $l$ converters; $\eta_{B1i}(j\omega_{c1}) = (\phi_{B1i}(j\omega_{c1}))^2/U_i^2$ represents the weighted coefficient before the power adjustment.

Then, ($k-m$) converters at active power generation states are adjusted into those with active power consumption states ($m<k$). That is, the positive inertia of the matrix $\tilde{P}$ decreases from $i_+(\tilde{P}) = k$ to $i_+(\tilde{P}) = m$. The proposed stability indicator of the MCPS can be changed as

$$D_{net1}^{ad} = -\sum_{i=1}^{m} \eta_{B1i}^{ad}(j\omega_{c1})P_{Gi} - \sum_{i=m+1}^{k-m} \eta_{B1i}^{ad}(j\omega_{c1})P_{adi} - \sum_{i=k+1}^{l} \eta_{B1i}^{ad}(j\omega_{c1})P_{Ci} \quad (33)$$

where $P_{adi}$ ($i=m+1,\ldots,k$) represents the active power of ($k-m$) converters adjusted from the generation state to the consumption state, and satisfies $P_{adi} < 0 < P_{Gi}$; $\eta_{B1i}^{ad}(j\omega_{c1}) = (\phi_{B1i}^{ad}(j\omega_{c1}))^2/U_i^2$ represents the weighted coefficient after the power adjustment; $U_i$ is considered almost unchanged before and after power adjustment, due to the weak coupling between active power and voltage amplitude in the power system [29].

The following proposition shows the cancellation effect between the active power generation and consumption on the system stability.

**Proposition IV.2.** *The positive inertia of active power matrix $i_+(\tilde{P})$ represents the number of converters at active power generation states, which has a negative impact on the critical network-side damping coefficient $D_{net1}$. When some converters with active power generation are adjusted into those with active power consumption, $i_+(\tilde{P})$ can be decreased and then $D_{net1}$ in (32) and (33) can be strictly increased, that is,*

$$D_{net1}^{ad} > D_{net1} \quad (34)$$

*Proof.* The detailed proof is given in Appendix D.

We remark that the negative impact on the stability indicator caused by converter-based active power generation can be mitigated to some extent by the effect of converter-based active power consumption. As a result, the cancellation between active power generation and consumption can provide insights into preventing SSI issues in the MCPS. These results potentially lay the foundation for understanding the interaction between converter-based generators and loads [2], and coordinating their operation in the future MCPS.

## V. CASE STUDIES

The validity of the proposed small-signal synchronization stability indicator and the correctness of the proposed role of multi-converter complex power will be verified in this section. Without loss of generality, a multi-converter power system consisting of WTGs and ESs is investigated as the test system as shown in Fig. 7. This system is modified from a realistic renewables energy station with high wind penetration in China. More specifically, there are three type-IV WTGs connected at Node 2, Node 4 and Node 5, and two ESs connected at Node 1 and Node 3. Node 6 represents the infinite bus equivalent to the external ac grid in Fig. 7. The WTG 1~3 can be controlled in the active power generation state and ES 1~2 can be controlled in the complex power generation and consumption states, that is, operating at four-quadrant states [9]. The parameters of converters of WTGs and ESs and parameters of networks are given in TABLE A.I- TABLE A.III in Appendix A.

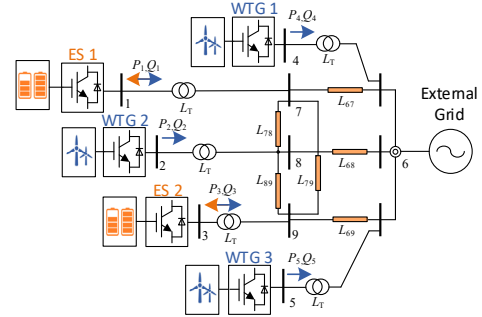

Fig. 7 A multi-converter power system consisting of WTGs and ESs.

*A. Validation of Proposed Stability Indicator*

To verify the efficacy of the proposed stability indicator, we consider three cases with different operating complex power of WTGs and ESs. The operating complex power of converters in Case 1~3 is presented in TABLE. I.

TABLE. I
OPERATING COMPLEX POWER OF CONVERTERS IN THREE CASES (PER-UNIT)

| Case | 1 | 2 | 3 |
|---|---|---|---|
| ES 1 | $P_1$= -0.86, $Q_1$= 0 | $P_1$= 0.8, $Q_1$= 0.60 | $P_1$= 1.0, $Q_1$= 0 |
| WTG 1 | $P_2$=0.4, $Q_2$= 0 | $P_2$= 0.78, $Q_2$=0.63 | $P_2$= 0.98, $Q_2$=0.15 |
| ES 2 | $P_3$= -0.6, $Q_3$= 0.2 | $P_3$= 0.6, $Q_3$= 0 | $P_3$= 1.0, $Q_3$= 0 |
| WTG 2 | $P_4$= 0.67, $Q_4$= 0.3 | $P_4$= 0.95, $Q_4$= 0.3 | $P_4$= 0.95, $Q_4$= 0.3 |
| WTG 3 | $P_5$= 0.70, $Q_5$= 0 | $P_5$= 0.86, $Q_5$= -0.5 | $P_5$= 0.86, $Q_5$= -0.5 |
| $D_{net1}$ | $-0.04>-D_{con}(\omega_{c1})$ | $-0.08=-D_{con}(\omega_{c1})$ | $-0.12<-D_{con}(\omega_{c1})$ |
| Stability | Stable | Unstable | Unstable |

Under three cases, the curves of converter-side and network-side damping coefficients and spring coefficients can be calculated, as shown in Fig. 8. Based on Fig. 8, the values of proposed stability indicator (i.e., $D_{net1}$) and the converter-side damping coefficient at oscillation angular frequency (i.e. $-D_{con}(\omega_{c1})$) under Case 1~3 can be obtained and also listed in TABLE. I. It indicates that $D_{net1}$ is determined by the operating complex power of WTGs and ESs. When WTGs and ESs operate at the complex power points given in Case 1, the damping coefficient relationship at the oscillation frequency ($f_{c1}$=20.0Hz) satisfies $D_{net1} > -D_{con}(\omega_{c1})$, thereby leading to a positive net damping and a stable operation. In contrast, when WTGs and ESs operate at the complex power injection given in Case 2 and Case 3, we have $D_{net1} \leq -D_{con}(\omega_{c1})$ at the oscillation frequency (both $f_{c1}$=20.3Hz in Case 2 and Case3 within the PLL bandwidth), thus indicating the SSI issues.

The efficacy of the proposed stability indicator is validated by comparing the results of TABLE. I and Fig. 8 with those of eigenvalue analysis and time-domain electromagnetic transient simulation in Fig. 9 and Fig. 10, respectively.

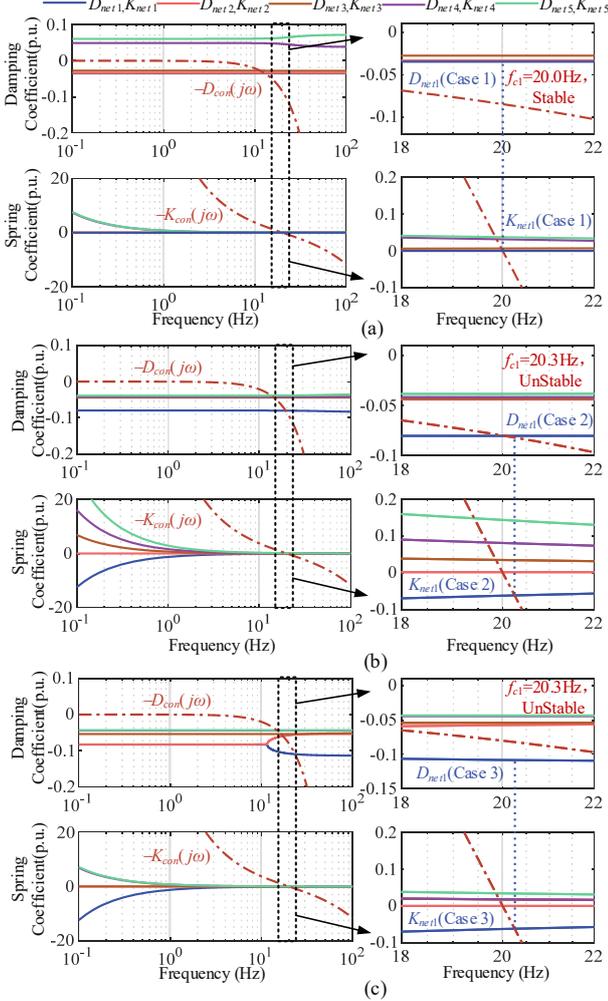

Fig. 8 Curves of converter-side and network-side damping coefficients and spring coefficients (a) in Case 1;(b) in Case 2;(c) in Case 3.

The eigenvalue analysis results in Fig. 9 are consistent with the results of TABLE. I and Fig. 8. Fig. 9 shows that the dominant eigenvalues in Case 1 are located in the left-half plane, which indicates stable operation. On the contrary, the dominant eigenvalues in Case 2~3 are located in the right-half plane, which means the instability oscillation risks. Also, the oscillation frequencies in Case 2~3 are demonstrated to be 20.4 Hz and 20.1 Hz, which are very close to the oscillation frequency ($f_{c1}$ = 20.3Hz) given in Fig. 8.

Moreover, results of TABLE. I and Fig. 8 are consistent with those of the time-domain electromagnetic transient simulation as depicted in Fig. 10. Under three cases, the same voltage surge 10% is applied at 2.0 s to Node 6 and then is cleared at 2.02s in the system. Fig. 10 (b) and (c) show that the divergent active power and reactive power oscillations occur in Case 2~3 following the disturbance; whereas such oscillations do not occur in Case 1 as shown in Fig. 10 (a). Thus, this consistency between the results of the proposed stability indicator and those of simulations verifies the effectiveness of the proposed stability indicator in the varied operating complex power.

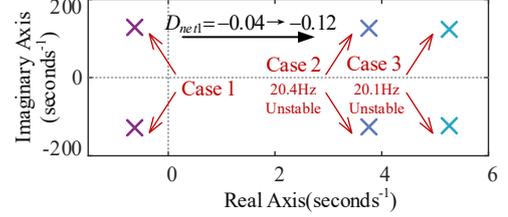

Fig. 9 Dominant eigenvalues of the test system in Case 1~3.

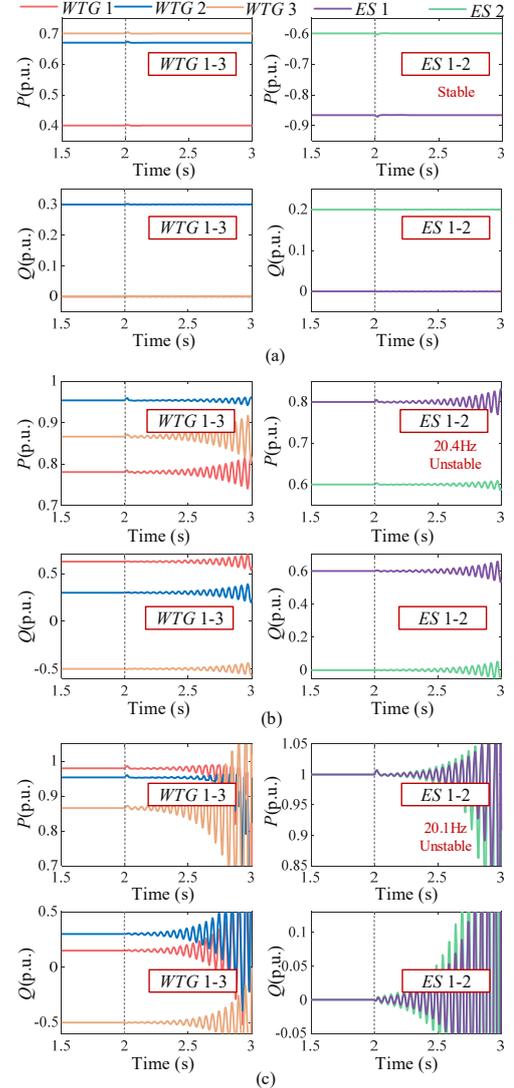

Fig. 10 Trajectories of active power and reactive power of WTGs and ESs in the test system (a) in Case 1;(b) in Case 2;(c) in Case 3.

*B. Validation of Impact Analysis of Complex Power on Small-Signal Synchronization Stability*

To facilitate the validation of our impact analysis results of complex power on the system stability, we now consider two scenarios with complex power variations based on Case 2 in TABLE. I.

*Scenario 1. Complex Power Variation Influence*: To verify the proposed role of complex power variation impact in Proposition IV.1, we successively increase the active power and reactive power of each WTG and ES converter by the same amplitude (i.e., $\Delta P_i$=0.1 p.u. and $\Delta Q_i$=0.1 p.u., $i=\{1,…,5\}$), and the other parameters remain unchanged. For each variation, the curves of converter-side and network-side damping coefficients and spring coefficients can be calculated, and the critical curves are shown in Fig. 11.

Fig. 11(a) shows the critical network-side damping and spring coefficients when the active power of each WTG and ES increases $\Delta P_i$=0.1 p.u., $i=\{1,…,5\}$, successively. It can be seen that the critical network-side damping coefficient is significantly decreased with the active power increasing, whereas the spring coefficients are almost unchanged, and thus lead to the deterioration of system stability.

Also, Fig. 11(b) shows the critical network-side damping and spring coefficients while the reactive power of each WTG and ES increases $\Delta Q_i$=0.1 p.u., $i=\{1,…,5\}$, successively. Compared with the impact of the same amplitude active power variations, it can be seen that the change of the critical network-side damping coefficients with reactive power variations are less sensitive.

results in Fig. 11(a), that is, the critical network-side damping coefficient $D_{net1}$ is most sensitive to the active power variation of WTG 1($\Delta P_2$=0.1 p.u). Thus, the above results in Scenario 1 are consistent with the proposed role in Proposition IV.1, which can verify its correctness.

*Scenario 2. Interactions Between Active Power Generation and Consumptions*: To verify the proposed role of interactions between active power generation and consumption in Proposition IV.2, the operating active power states of ES 1~2 in Case 2 are adjusted from generation to consumption (i.e., $P_1$ = 0.8 p.u. to -0.8 p.u. and $P_3$ = 0.6 p.u. to -0.6 p.u.), and operating power points of WTGs remain unchanged.

Fig. 12 shows the critical network-side damping and spring coefficients before and after the adjustment. It can be seen that the spring coefficients of networks are almost unchanged, whereas the critical network-side damping coefficient significantly increases after the active power adjustment. Also, we have $D_{net1} > -D_{con}(\omega_{c1})$ after the active power adjustment and it means the small-signal synchronization stability enhancement.

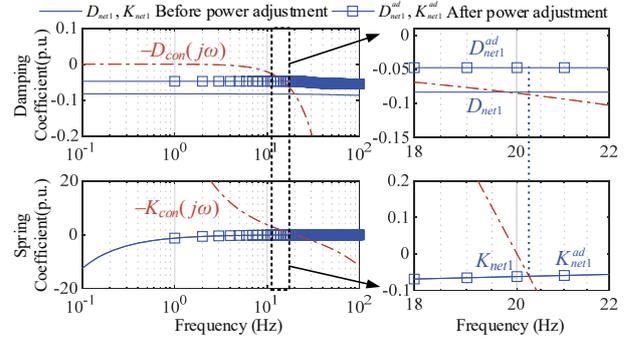

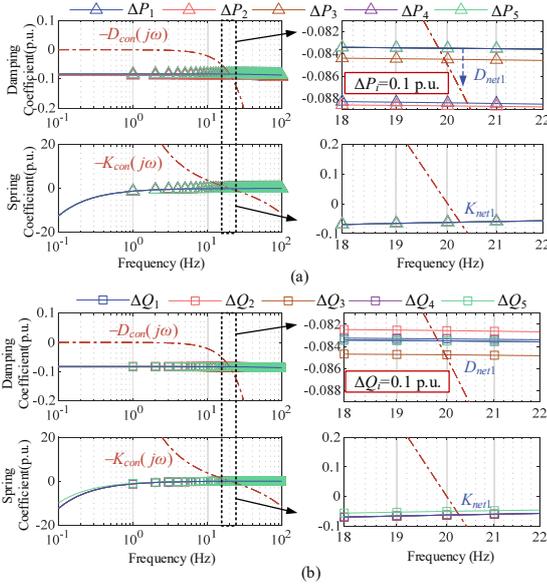

Fig. 11 Curves of critical damping and spring coefficients of testing system with the complex power of converters variations. (a) increase $\Delta P$=0.1 p.u.;(b) increase $\Delta Q$=0.1 p.u..

TABLE. II
WEIGHTED COEFFICIENT RESULTS FOR ACTIVE POWER INJECTION ADJUSTMENT

| Converter | ES1 | WTG1 | ES2 | WTG2 | WTG3 |
|---|---|---|---|---|---|
| Weighted Coefficient $\|\eta_{B1i}(j\omega_{c1})\|$ | 0.047 | **0.051** | 0.011 | 1e-6 | 1e-6 |

Moreover, the norm of weighted coefficients $|\eta_{B1i}(j\omega_{c1})|$ ($i=\{1,…,5\}$) of the proposed stability indicator in Case 2 can be calculated and given in TABLE. II. It can be seen that WTG 1 can be identified as the dominant converter since it has the largest weighted coefficient. This result is consistent with the

Fig. 12 Curves of critical converter-side and network-side damping coefficients and spring coefficients when adjusting active power generation to consumption.

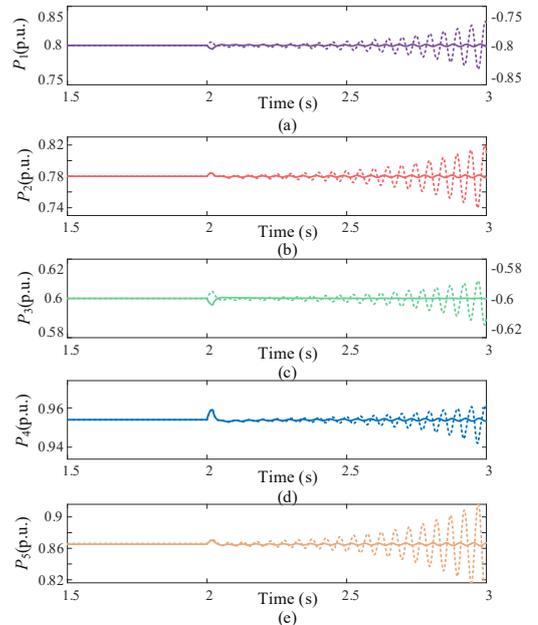

Fig. 13 Time-domain simulation results in the testing system when adjusting active power generation to consumption (dashed line represents results before adjustment and solid line represents results after adjustment). (a) and (c) are the active powers of ES 1-2; (b), (d) and (e) are the active powers of WTG1-3.

Moreover, results in Fig. 12 are consistent with those of the time-domain electromagnetic transient simulation as depicted in Fig. 13. Both in the systems before and after the active power adjustment, the same voltage surge 10% is applied at 2.0 s to Node 6 and then is cleared at 2.02s in the system. It can be seen that the divergent power oscillations occurred before the power adjustment in Case 2 are suppressed after the power adjustment, and means stable operation. Thus, this consistency between the results of analysis in Fig. 12 and those of simulations verifies the effectiveness of the proposed role in Proposition IV.2.

## VI. CONCLUSION

This paper investigated the role of complex power in the small-signal synchronization stability of MCPS. We explicitly showed how PLL dynamics of converters in the MCPS are coupled via the power network and how analogously apply the positive-net-damping criterion to analyze the small-signal synchronization stability of MPCS. Based on these insights, a systematical stability indicator was proposed to quantify the stability margin under various complex power points. Moreover, we provided a weighted summation form of the proposed stability indicator, which facilitated to reveal the role of each converter's complex power variations in the system stability though sensitivity analysis. We proved that the impact of active power variation on the system stability is more sensitive than that of reactive power variation in the MCPS and showed increasing active power injection of the dominant converter can significantly deteriorate the system stability. Further, we revealed the cancellation effect between the active power generation and consumption, which is particularly useful for preventing SSI issues in the MCPS. The correctness and effectiveness of our theoretical findings were confirmed by the simulations results. The future work is developing the coordination strategies for complex power generation and consumption of converter-based apparatuses to improve the system stability optimally.

## APPENDIX

### A. System Parameters

TABLE A.I
PARAMETERS OF CONVERTER OF WIND TURBINE GENERATOR

| Apparatus Base Values for Per-unit Calculation |
|---|
| $f_{base}$=50Hz   $U_{base}$=0.69 kV   $S_{base}$= 1.5MVA |
| Power Factor Range |
| 0.8 Leading to 0.8 Lagging |
| Parameters of Control Loops (per-unit values) |
| PI parameters of the current control Loop: 8, 40 |
| Time coefficient of power measurement: 0.01 |
| PI parameters of the active power control Loop: 0.3, 10 |
| PI parameters of the reactive power control Loop: 0.3, 10 |
| PI parameters of the PLL: 6.5, 15782 (PLL bandwidth: 30.2 Hz) |

TABLE A.II
PARAMETERS OF CONVERTER OF ENERGY STORAGE

| Apparatus Base Values for Per-unit Calculation |
|---|
| $f_{base}$=50Hz   $U_{base}$=0.69 kV   $S_{base}$= 1.5MVA |
| Power Factor Range |
| -1.0 to 1.0 |
| Parameters of Control Loops (per-unit values) |
| PI parameters of the current control Loop: 0.3, 10 |
| Time coefficient of power measurement: 0.01 |
| PI parameters of the active power control Loop: 0.5, 40 |
| PI parameters of the reactive power control Loop: 0.5, 40 |
| PI parameters of the PLL: 6.5, 15782 (PLL bandwidth: 30.2 Hz) |

TABLE A.III
NETWORK PARAMETERS OF TEST SYSTEM IN PER UNIT

| $L_T$ = 0.05 | $L_{68}$ = 0.0725 | $L_{78}$ = 0.0145 |
|---|---|---|
| $L_{67}$ = 0.0667 | $L_{69}$ = 0.0696 | $L_{89}$ = 0.0232 |

### B. Proof of Lemma 1

Since matrix $\boldsymbol{B}$ is positive definite, the matrix $\boldsymbol{G}_{net}(j\omega)$ in (19) satisfies $\boldsymbol{B}^{-1/2}\boldsymbol{G}_{net}(j\omega)\boldsymbol{B}^{1/2} = \boldsymbol{G}'_{net}(j\omega)$. This concludes the proof ∎.

### C. Proof of Lemma 2

According to Lemma 1, the eigenvalue $\lambda_1\{\boldsymbol{G}'_{net}(j\omega_{c1})\}$ can be used to characterize the properties of $\lambda_1\{\boldsymbol{G}_{net}(j\omega_{c1})\}$ and we have

$$\boldsymbol{G}'_{net}(j\omega_{c1})\phi_1 = \lambda_1\phi_1 \Rightarrow \begin{cases} \lambda_1 = \phi_1^* \boldsymbol{G}'_{net}(j\omega_{c1})\phi_1 \\ \overline{\lambda}_1 = \phi_1^* \boldsymbol{G}'^*_{net}(j\omega_{c1})\phi_1 \end{cases} \quad (35)$$

where $\phi_1(j\omega_{c1}) \in \mathbb{C}^n$ represents the normalized right eigenvector with respect to the eigenvalue $\lambda_1\{\boldsymbol{G}'_{net}(j\omega_{c1})\}$, which satisfies $\phi_1^*\phi_1 = 1$.

Also, according to [30], $\boldsymbol{G}_{net}(j\omega_{c1})$ in (27) has the Cartesian(or Toeplitz) decomposition, that is,

$$\begin{cases} -\boldsymbol{B}^{-1/2}\tilde{\boldsymbol{P}}\boldsymbol{B}^{-1/2} = \dfrac{1}{2}(\boldsymbol{G}'_{net}(j\omega_{c1}) + \boldsymbol{G}'^*_{net}(j\omega_{c1})) \\ \omega_{R1}\boldsymbol{B}^{-1/2}\tilde{\boldsymbol{Q}}\boldsymbol{B}^{-1/2} = \dfrac{1}{2j}(\boldsymbol{G}'_{net}(j\omega_{c1}) - \boldsymbol{G}'^*_{net}(j\omega_{c1})) \end{cases} \quad (36)$$

Combing (36) and (35), the real part and imaginary part of eigenvalue $\lambda_1\{\boldsymbol{G}'_{net}(j\omega_{c1})\}$ (equivalent to $\lambda_1\{\boldsymbol{G}_{net}(j\omega_{c1})\}$) can be written as

$$\begin{cases} \operatorname{Re}\lambda_1\{\boldsymbol{G}_{net}(j\omega_{c1})\} = \dfrac{\lambda_1+\overline{\lambda}_1}{2} = \phi_1^* \dfrac{\boldsymbol{G}'_{net}+\boldsymbol{G}'^*_{net}}{2}\phi_1 = -\phi_1^*\boldsymbol{B}^{-1/2}\tilde{\boldsymbol{P}}\boldsymbol{B}^{-1/2}\phi_1 \\ \operatorname{Im}\lambda_1\{\boldsymbol{G}_{net}(j\omega_{c1})\} = \dfrac{\lambda_1-\overline{\lambda}_1}{2j} = \phi_1^* \dfrac{\boldsymbol{G}'_{net}-\boldsymbol{G}'^*_{net}}{2j}\phi_1 = \omega_{R1}\phi_1^*\boldsymbol{B}^{-1/2}\tilde{\boldsymbol{Q}}\boldsymbol{B}^{-1/2}\phi_1 \end{cases}$$
(37)

Substituting $\phi_{B1}(j\omega_{c1}) = \boldsymbol{B}^{-1/2}\phi_1(j\omega_{c1})$ in (37) yields

$$\begin{cases} \operatorname{Re}\lambda_1\{\boldsymbol{G}_{net}(j\omega_{c1})\} = -\phi_{B1}^*(j\omega_{c1})\tilde{\boldsymbol{P}}\phi_{B1}(j\omega_{c1}) \\ \operatorname{Im}\lambda_1\{\boldsymbol{G}_{net}(j\omega_{c1})\} = \omega_R\phi_{B1}^*(j\omega_{c1})\tilde{\boldsymbol{Q}}\phi_{B1}(j\omega_{c1}) \end{cases} \quad (38)$$

Then, substitute diagonal matrices $\tilde{\boldsymbol{P}} = diag\{P_i/U_i^2\}$ and $\tilde{\boldsymbol{Q}} = diag\{Q_i/U_i^2\}$ into (38), and let $\eta_{B1i}(j\omega_{c1}) = \phi_{B1i}^2(j\omega_{c1})/U_i^2$. This concludes the proof ∎.

### D. Proof of Proposition IV.2

By considering that $\phi_{B1}$ in (32) and $\phi_{B1}^{ad}$ in (33) both satisfy $\phi_{B1}^*\boldsymbol{B}\phi_{B1} = 1$ and $\phi_{B1}^{ad*}\boldsymbol{B}\phi_{B1}^{ad} = 1$, the norm of these vectors remains unchanged before and after the power adjustment. Then, existing a unitary matrix $\boldsymbol{R} \in \mathbb{C}^{n \times n}$ makes

$$\phi_{B1}^{ad} = \boldsymbol{R}\phi_{B1}, \boldsymbol{R}^*\boldsymbol{R} = \boldsymbol{I} \quad (39)$$

Substituting (39) into (32) and (33) yields

$$\begin{cases} D_{net1}^{ad} = -\phi_{B1}^{ad*}\tilde{\boldsymbol{P}}^{ad}\phi_{B1}^{ad} = -\phi_{B1}^*\boldsymbol{R}^*\tilde{\boldsymbol{P}}^{ad}\boldsymbol{R}\phi_{B1}, \\ D_{net1} = -\phi_{B1}^*\tilde{\boldsymbol{P}}\phi_{B1} \end{cases} \quad (40)$$

where $\tilde{\boldsymbol{P}}^{ad}$ represents the active power matrix after the power adjustment whose elements are $P_{Gi}$, $P_{adi}$, and $P_{Ci}$ in (33).

Since the active power of ($m$+1)-th to $k$-th converters are adjusted from the generation to the consumption and satisfy $P_{adi} < 0 < P_{Gi}$ ($i=m+1,…,k$), we have $i_+(\tilde{\boldsymbol{P}}^{ad}) = m < i_+(\tilde{\boldsymbol{P}}) = k$ and $\tilde{\boldsymbol{P}}^{ad} < \tilde{\boldsymbol{P}}$. And consider that the matrices $\tilde{\boldsymbol{P}}$ and $\tilde{\boldsymbol{P}}^{ad}$ in (40) are Hermitian [30] and we have $\boldsymbol{R}^*\tilde{\boldsymbol{P}}^{ad}\boldsymbol{R} < \tilde{\boldsymbol{P}}$.

Then, according to Theorem 8.11 in [31], for the vector $\phi_{B1}(j\omega_{c1}) \in \mathbb{C}^n$ in (40), we have

$$-\phi_{B1}^*(j\omega_{c1})\boldsymbol{R}^*\tilde{\boldsymbol{P}}^{ad}\boldsymbol{R}\phi_{B1}(j\omega_{c1}) > -\phi_{B1}^*(j\omega_{c1})\tilde{\boldsymbol{P}}\phi_{B1}(j\omega_{c1}) \quad (41)$$

Thus, we have $D_{net1}^{ad} > D_{net1}$ and this concludes the proof ∎.

**Fuyilong Ma** received the B.Eng. degree in electrical engineering from the College of Electrical Engineering, Zhejiang University, Hangzhou, China, in 2019. He is currently working toward the Ph.D. degree in the College of Electrical Engineering, Zhejiang University, Hangzhou. His research interests include renewable energy stability analysis and control.

**Huanhai Xin** (M'14) received the Ph.D. degree in College of Electrical Engineering, Zhejiang University, Hangzhou, China, in June 2007. He was a post-doctor in the Electrical Engineering and Computer Science Department of the University of Central Florida, Orlando, from June 2009 to July 2010. He is currently a Professor in the Department of Electrical Engineering, Zhejiang University. His research interests include distributed control in active distribution grid and micro-gird, AC/DC power system transient stability analysis and control, and grid-integration of large-scale renewable energy to weak grid.

**Zhiyi Li** (GSM'14-M'17) received the Ph.D. degree in Electrical Engineering from Illinois Institute of Technology in 2017. He received M.E. degree in Electrical Engineering from Zhejiang University (Hangzhou, China) in 2014 and the B.E. degree in Electrical Engineering from Xi'an Jiaotong University (Xi'an, China) in 2011. From August 2017 to May 2019, he was a senior research associate at Robert W. Galvin Center for Electricity Innovation at Illinois Institute of Technology. Since June 2019, he has been with the College of Electrical Engineering, Zhejiang University (Hangzhou, China) as a research professor. His research interests lie in the application of state-of-the-art optimization and control techniques in smart grid design, operation and management with a focus on cyber-physical security.

**Linbin Huang** (Member, IEEE) received the B.Eng. and Ph.D. degrees from Zhejiang University, Hangzhou, China, in 2015 and 2020, respectively. He is currently a Post-Doctoral Researcher with the Automatic Control Laboratory, ETH Zürich, Zürich, Switzerland. His research interests include power system stability, optimal control of power electronics, and data-driven control